\documentclass[twocolumn]{aastex61}

\usepackage{epsfig}
\usepackage{epstopdf}

\newcommand{\lapp}{$_<\atop{^\sim}$}
\newcommand{\gapp}{$_>\atop{^\sim}$} 

\shortauthors{Rhode et al.}
\shorttitle{Properties of Andromeda Satellite Dwarf Lac~I}

\begin{document}

\title{Structural and Photometric Properties of the Andromeda
  Satellite Dwarf Galaxy Lacerta~I from Deep Imaging with WIYN pODI}

\author{Katherine L. Rhode}
\affiliation{Indiana University Department of
    Astronomy, 727 East 3rd Street, Swain West 319, Bloomington, IN
    47405, USA} 
\email{krhode@indiana.edu}

\author{Denija Crnojevi\'c}
\affil{Texas Tech University Department of Physics, Box
  41051, Lubbock, TX 79409-1051, USA} 
\email{denija.crnojevic@ttu.edu}

\author{David J. Sand}
\affil{Texas Tech University Department of Physics, Box
  41051, Lubbock, TX 79409-1051, USA} 
\email{david.sand@ttu.edu}

\author{Steven Janowiecki} 
\affil{International Centre for Radio
  Astronomy Research (ICRAR), University of Western Australia, 35
  Stirling Highway, Crawley, WA 6009, Australia}
\email{steven.janowiecki@uwa.edu.au}

\author{Michael D. Young}
\affiliation{Indiana University Department of Astronomy, 727 East 3rd Street,
Swain West 319, Bloomington, IN 47405, USA}
\email{youngmd@indiana.edu}

\author{Kristine Spekkens}
\affiliation{Department of Physics, Royal Military College of
  Canada, P.O. Box 17000, Station Forces, Kingston, Ontario, K7K 7B4,
  Canada}
\email{Kristine.Spekkens@rmc.ca}

\begin{abstract}
We present results from WIYN pODI imaging of Lacerta~I (And XXXI), a
satellite dwarf galaxy discovered in the outskirts of the Andromeda
galaxy (M31) in Pan-STARRS1 survey data.  Our deep, wide-field $g,i$
photometry reaches $\sim$3~magnitudes fainter than the photometry in
the Pan-STARRS1 discovery paper and allows us to trace the stellar
population of Lac~I beyond two half-light radii from the galaxy
center.  We measure a Tip of the Red Giant Branch (TRGB) distance for
Lac~I of $(m-M)_0=24.44\pm0.11$ mag (773$\pm$40~kpc, or 264$\pm$6~kpc
from M31), which is consistent with the Pan-STARRS1 distance. We use a
maximum-likelihood technique to derive structural properties for the
galaxy, and find a half-light radius ($r_h$) of 3.24$\pm$0.21~arcmin
(728$\pm$47~pc), ellipticity ($\epsilon$) of 0.44$\pm$0.03, total
magnitude $M_V$ $=$ $-$11.4$\pm$0.3, and central surface brightness
$\mu_{V,0}$ $=$ 24.8$\pm$0.3 mag~arcsec$^{-2}$.  We find no HI
emission in archival data and set a limit on Lac~I's neutral gas
mass-to-light ratio of $M_{HI}/L_V$ $<$ 0.06 $M_\sun$/$L_\sun$,
confirming Lac~I as a gas-poor dwarf spheroidal galaxy.  Photometric
metallicities derived from Red Giant Branch stars within 2~$r_h$ yield
a median [Fe/H] of $-$1.68$\pm$0.03, which is more metal-rich than the
spectroscopically-derived value from Martin et al. (2014). Combining
our measured magnitude with this higher metallicity estimate places
Lac~I closer to its expected position on the luminosity-metallicity
relation for dwarf galaxies. 
\end{abstract}

\keywords{Local Group -- galaxies: individual (Lac~I) -- galaxies: individual (And~XXXI) -- galaxies: dwarf -- galaxies: photometry}

\section{Introduction}
\label{section: introduction}

The number, spatial distribution, mass function, and kinematics of the
dwarf galaxies in an environment like that of the Local Group provide
important tests for theoretical ideas about cosmology, dark matter,
and galaxy formation (e.g., Klypin et al. 1999, Moore et al. 1999,
Bullock \& Johnston 2005, Simon \& Geha 2007).  Furthermore, dwarf
galaxies themselves serve as valuable laboratories for our
understanding of the physical processes involved in galaxy evolution,
star formation, feedback, and chemical evolution (e.g., Mateo 1998,
Tolstoy et al. 2009. McConnachie 2012 and references therein).  The
importance of dwarf galaxies to all of these areas of extragalactic
astrophysics and cosmology has motivated an array of surveys and
searches for additional low-mass galaxies
%in and around the Local Group (e.g., Willman et al. 2005, Martin et
%al. 2006, Belokurov et al. 2007, Bell et al. 2011, Adams et al. 2013).
in and around the Local Group (e.g., Willman et al. 2005; Martin et
al. 2006; Belokurov et al. 2007; Bell et al. 2011; McConnachie et
al. 2009; Adams et al. 2013; Bechtol et al. 2015; Koposov et al. 2015;
Kim et al. 2015, Janesh et al. 2015).

In particular, the regions around the Andromeda Galaxy (M31) have
lately been the focus of a number of dedicated searches for satellite
dwarf galaxies, and have met with much success.  For example, searches
for stellar overdensities in photometric observations from the Sloan
Digital Sky Survey (SDSS) and the Pan-Andromeda Archaeological Survey
(PAndAS) have resulted in the discovery of tens of dwarf galaxies in
regions around Andromeda (e.g., Zucker 2004, 2007; McConnachie et
al. 2008; Martin et al. 2006, 2009; Bell et al. 2011; Slater et
al. 2011). A recent contribution in this area comes from the Panoramic
Survey Telescope and Rapid Response System 1 (Pan-STARRS1; Kaiser et
al. 2010) 3$\pi$ survey.  By searching through Pan-STARRS1 photometric
source catalogs, Martin et al. (2013a) discovered two new dwarf
galaxies, Lacerta~I (And XXXI) and Cassiopeia~III (And XXXII), in
regions around Andromeda that had not been included in other
systematic imaging surveys.  In a follow-up paper, Martin et
al. (2013b) described the discovery in Pan-STARRS1 imaging data
of a third Andromeda satellite galaxy, Perseus~I (And XXXIII), located
in a region with shallow SDSS coverage. All three galaxies were
confirmed to be satellite galaxies via a spectroscopic study by
\citet{martin14}, who derived systemic radial velocities for each
system as well as individual metallicity measurements for member stars
(see Section~\ref{section: metallicity} for more discussion).  All
three
%of the galaxies 
%These three newly-discovered dwarf galaxies
%Lac~I (And XXXI), Cas III (And XXXII), and Per I (And XXXIII), 
are located more than 10 degrees away from M31 in projection and have
relatively faint central surface brightnesses ($\mu_0$ \gapp 25$-$26
mag~arcsec$^{-2}$),
%Per I 25.7, Lac I 25.8 and Cas III 26.4 in discovery paper
properties which likely contributed to their late discovery (Martin et
al.\ 2013a, 2013b). Cas~III is located within the area of the thin
rotating plane of M31 satellite galaxies \citep{ibata13} but
\citet{martin14} showed that it is moving in the opposite sense
relative to the disk; Per~I and Lac~I are located on the far eastern
and far western sides of Andromeda, respectively
\citep{martin13a,martin13b}.

Prompted in part by the installation and commissioning of a new camera
on the WIYN 3.5-m telescope\footnote{The WIYN Observatory is a joint
  facility of the University of Wisconsin-Madison, Indiana University,
  the National Optical Astronomy Observatory and the University of
  Missouri.} during the 2012$-$2013 observing season, we began a
campaign to obtain deep, wide-field imaging of these and other
selected nearby dwarf galaxies.  The aim is to take advantage of the
excellent image quality and depth made possible by WIYN to study the
galaxies' structure and stellar populations out to large
galactocentric radius.  The first dwarf galaxy we targeted is Lac~I, a
relatively luminous dwarf ($M_V$ $\sim$ $-$12) that lies
$\sim$20$^\circ$ ($\sim$275~kpc) away from Andromeda in projected
distance (Martin et al.\ 2013a).  The photometry of Lac~I presented in
Martin et al. (2013a) reached $i$ $\sim$ 22.5 and yielded estimates of
the distance, size, and metallicity of the galaxy.
%($r_h$ $=$ 4.2$^{+0.4)_{-0.5}$$\arcmin$), 
The \citet{martin14} spectroscopy study presented the systemic
velocity and velocity dispersion
($v_{r,helio}$$=$$-$198.4$\pm$1.4~km~s$^{-1}$,
$\sigma_{v,r}$$=$10.3$\pm$0.9~km~s$^{-1}$) as well as a refined
metallicity estimate and $V$-band mass-to-light ratio. In this paper,
we present results from imaging of Lac~I that reaches $\sim$24$-$25 in
the $g$ and $i$ filters, covers a $\sim$20$\arcmin$ x 20$\arcmin$
area, and allows us to trace the stellar population beyond two
half-light radii.

The paper is organized as follows. Section~\ref{section: observations}
describes the observations and initial data reduction and
Section~\ref{section: detection} discusses our methods for source
detection, photometry, and completeness testing.
Section~\ref{section: properties} presents the properties we measure
for Lac~I, including the color-magnitude diagram (CMD), distance via
the Tip of the Red Giant Branch (TRGB) method, structural parameters,
luminosity, limits on the neutral gas content, and the
photometrically-derived metallicity distribution function.  The last
section of the paper gives a summary and our final conclusions.

\section{Observations and Data Reduction}
\label{section: observations}

Observations of Lac~I were obtained on 2013 October 1 with the WIYN
3.5-m telescope and the One Degree Imager with a partially-filled
focal plane (pODI; Harbeck et al. 2014). The pODI camera was comprised
of nine orthogonal transfer arrays (OTAs) arranged in a 3x3
configuration, as well as four additional OTAs positioned at various
radial locations around
%rest of 
the focal plane. Each individual OTA is an 8x8 arrangement of
orthogonal transfer CCD detectors with 480x496 12-$\micron$ pixels.
The central 3x3 array of OTAs in pODI, which provided a field-of-view
of $\sim$24$\arcmin$ x 24$\arcmin$ and a pixel scale of
0.11$\arcsec$~pixel$^{-1}$, was used to image the target objects and
the outlying OTAs were used for guiding during the exposure.
%The central array has a field-of-view of 24$\arcmin$ x 24$\arcmin$
%and a pixel scale of 0.11$\arcsec$~pixel${-1}$. 
(Note that pODI was upgraded in 2015 and is now referred to as the ODI
camera; new detectors were added to create a 5x6 OTA configuration
that provides a 40$'$ x 48$'$ field-of-view.)  We obtained nine
700-second exposures of the Lac~I field in $g$ and another nine
600-second exposures in $i$ with pODI.
%with the $g$-band filter and nine 600-second exposures with the
%$i$-band filter.
The telescope was dithered between exposures in order to help
eliminate gaps between the CCDs and OTAs during the subsequent image
stacking process.

The pODI images were immediately transferred from WIYN to the ODI
Pipeline, Portal, and Archive (ODI-PPA; Gopu et al. 2014)\footnote{The
  ODI Portal, Pipeline, and Archive (ODI-PPA) system is a joint
  development project of the WIYN Consortium, Inc., in partnership
  with Indiana University's Pervasive Technology Institute (PTI) and
  with the National Optical Astronomy Observatory Science Data
  Management (NOAO SDM) Program.} at Indiana University and later
processed with the QuickReduce pipeline \citep{kotulla14}. The
pipeline corrects each image for cross-talk, subtracts the overscan
signal, corrects the images for non-linearity and persistence, applies
bias, dark, flat-field, and pupil ghost corrections, and removes
cosmic rays.  The pipeline-processed images of Lac~I were then
flattened with a night-sky flat, reprojected to a common pixel scale,
and finally average-combined to create a single deep, stacked image in
each filter. Areas on the edges of the stacked images with slightly
higher noise levels (because of reduced exposure time due to the
dither pattern) were clipped to produce a usable field
$\sim$20$\arcmin$ x 20$\arcmin$ in size.  The mean full-width at
half-maximum of the point spread function (FWHM PSF) in the final
combined images is 0.84$\arcsec$ in the $g$-band and 0.70$\arcsec$ in
the $i$-band.

Sky conditions on the night that Lac~I was observed were clear.
Sloan Digital Sky Survey (SDSS; Ahn et al. 2012)
%KLR: I'm citing Ahn+2012 b/c we used DR9 mags for these steps.
stars present in various images of other fields taken throughout the
night were used to calculate photometric calibration coefficients that
could be applied to the Lac~I frames. (As mentioned in the
Introduction, the Lac~I field is not included in the current SDSS
footprint.)
% photometrically calibrate the images of Lac~I.  
The RMS scatter of the zero points calculated from individual SDSS
stars within a given image taken that night ranged from 0.016$-$0.021
magnitude.
%errors on the zero points for the photometric calibration were 
The {\it mean} zero points for images taken of different fields and at
different times of the night agreed with each other within $<$0.01
magnitude, confirming that the sky conditions were stable and the
night was photometric.  The photometric calibration coefficients
calculated in this way were applied to all instrumental magnitudes
measured via PSF photometry of the Lac~I images (see next section).
In addition, individual reddening corrections were calculated for each
star in the Lac~I images by applying the \citet{schlafly11}
coefficients to the \citet{schlegel98} values for Galactic extinction
at the position of the star.  The median color excess value across the
Lac~I field-of-view is $E(B-V)\sim0.14$.  For the remainder of the
paper,
%Throughout this paper, 
we will use final calibrated and dereddened $g_0$ and $i_0$ values.

\section{Source Detection and PSF Photometry}
\label{section: detection}

\begin{figure}
\centering
\includegraphics[width=8.5cm]{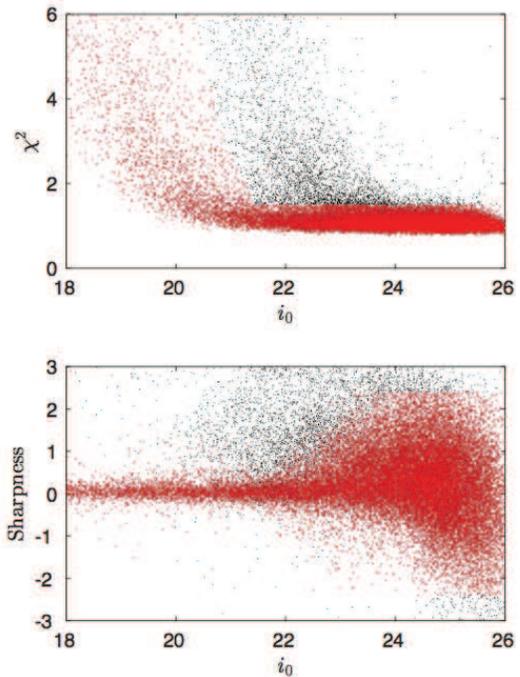}
\caption{Source selection criteria based on the $\chi^2$ and sharpness
  parameters from the DAOPHOT photometry, as a function of dereddened
  $i$-band magnitude. Red sources are the ones that we retain as
  genuine stars, black ones are rejected.}
\label{fig:cuts}
\end{figure}

Point-spread-function-(PSF)-fitting photometry was performed on the
final combined $g-$ and $i-$band images, using the suite of dedicated
programs DAOPHOT and ALLFRAME \citep{stetson87, stetson94}. We began
by selecting $\sim300$ bright, non-saturated stars in each image and
using them to construct a model PSF. This model PSF was then used to
measure magnitudes of all sources detected in each image with peak
counts at least $3\sigma$ above the background noise level.  The final
photometry catalog was cleaned by retaining sources with $\chi^2<1.5$
(i.e., the goodness-of-fit parameter for the PSF fitting of each
source) for $i_o>21.5$, and with a more generous cut below this value;
the sharpness was additionally constrained to be
$\left|sharp\right|<2$ (to remove, e.g., cosmic rays or resolved
background galaxies).  These cuts yield a total of $33577$ bona-fide
stars (see Figure~\ref{fig:cuts}).  The full photometric catalog for
the Lac~I $g$ and $i$ images is presented in
Table~\ref{tab:photometry}.  The catalog includes a sequence number
for each star, the star's Right Ascension and Declination, the
calibrated, dereddened $g_0$ and $i_0$ magnitudes and associated
instrumental errors, and the calculated Galactic extinction values
($A_g$ and $A_i$) that were applied.

In order to assess the true photometric uncertainties and
incompleteness in the images, we injected $\sim650,000$ artificial
stars (divided into 26 separate experiments to avoid artificial
crowding) distributed evenly across the frames and with colors and
magnitudes in the range covered by real stars.  We then performed the
same detection, photometry, calibration, and dereddening steps
described above to measure the magnitudes of the fake stars.  The
overall color-averaged $50\%$ completeness limit is reached at
$g_0=25.6$ and $i_0=24.2$, respectively. These values change to
$g_0=25.3$ and $i_i=24.0$ in the innermost $\sim3$~arcmin of Lac~I due
to higher stellar density there.  The artificial star tests indicate
that the photometric uncertainties are $\sim0.1$~mag at $i_0\sim23$
and $g_0\sim23.5$.

\section{Properties of Lac~I}
\label{section: properties}

\subsection{Color-Magnitude Diagram and Stellar Spatial Distribution}
\label{section: CMD}

\begin{figure*}
\centering
\includegraphics[width=16cm]{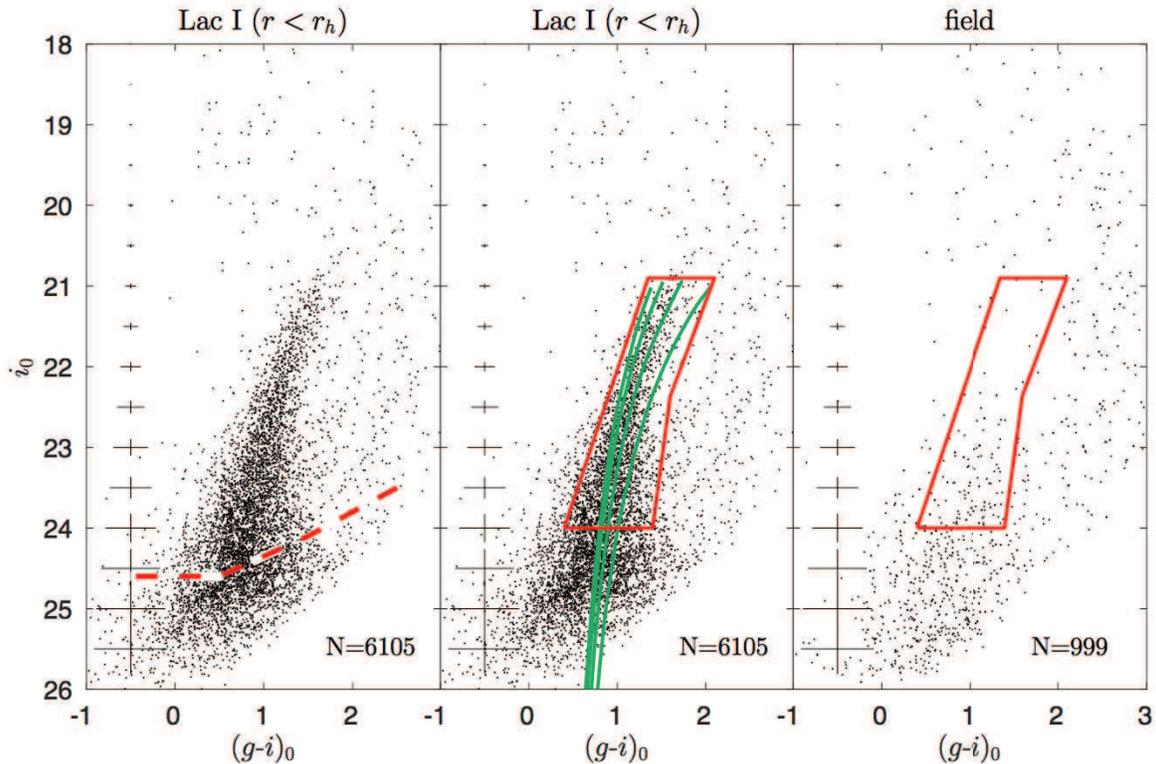}
\caption{Dereddened CMDs of all stars within Lac~I's half-light radius
  (\emph{left and middle panels}). The red dashed line in the left
  panel indicates the $50\%$ completeness level, and the photometric
  uncertainties derived from artificial star tests are shown in all
  the panels. The numbers of sources are also reported in each panel.
%  The area within the half-light radius
%  and the number of sources are also reported. 
  In the middle panel, 12~Gyr Dartmouth isochrones for 
  [Fe/H] = $-$2.5, $-$2.0, $-$1.5, and
  $-$1.0 (green lines) from \citet{dotter08} are overlaid on the data
  points, and the RGB selection box is marked (red polygon). The
  \emph{right panel} shows the combined CMD of the two field regions
  marked in Fig.~\ref{fig:spat}, rescaled to the area in the
  left and middle panels.}
%which cover an area $\sim2.8$ as
%  large as the area used to construct the Lac~I CMDs.  }
\label{fig:CMD}
\end{figure*}

\begin{figure*}
\centering
\includegraphics[width=3.5in]{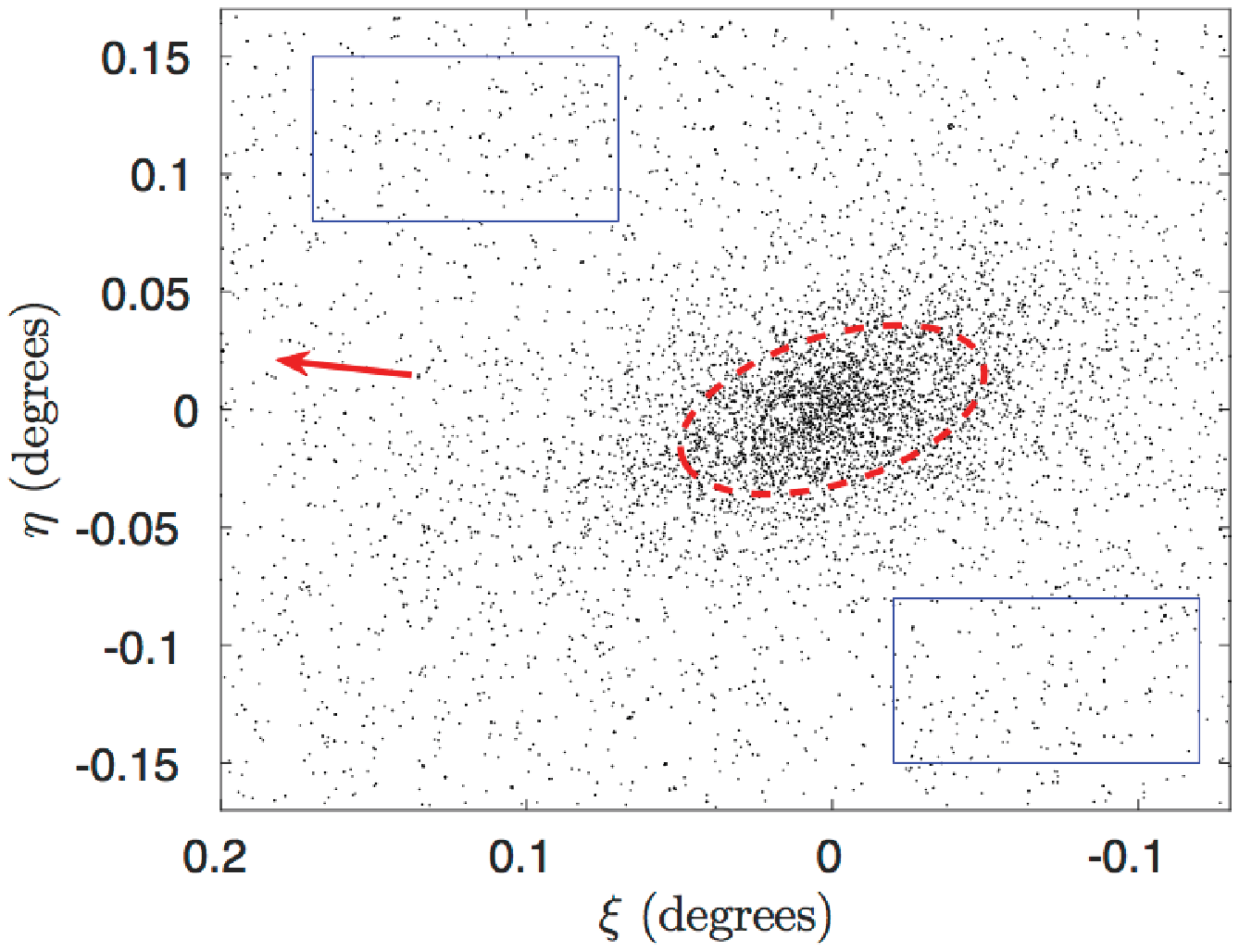}
\includegraphics[width=3.5in]{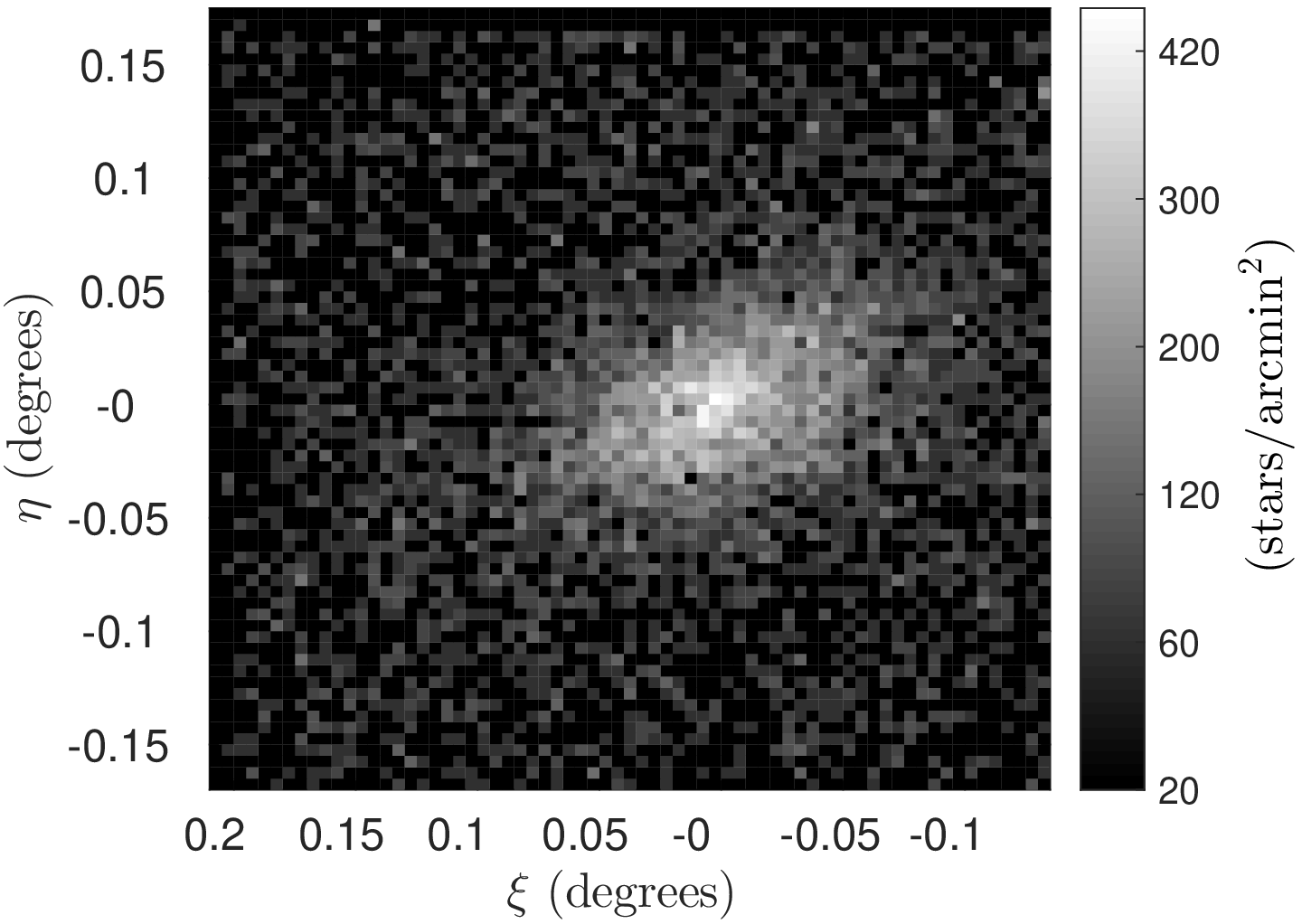}
\includegraphics[width=3.5in]{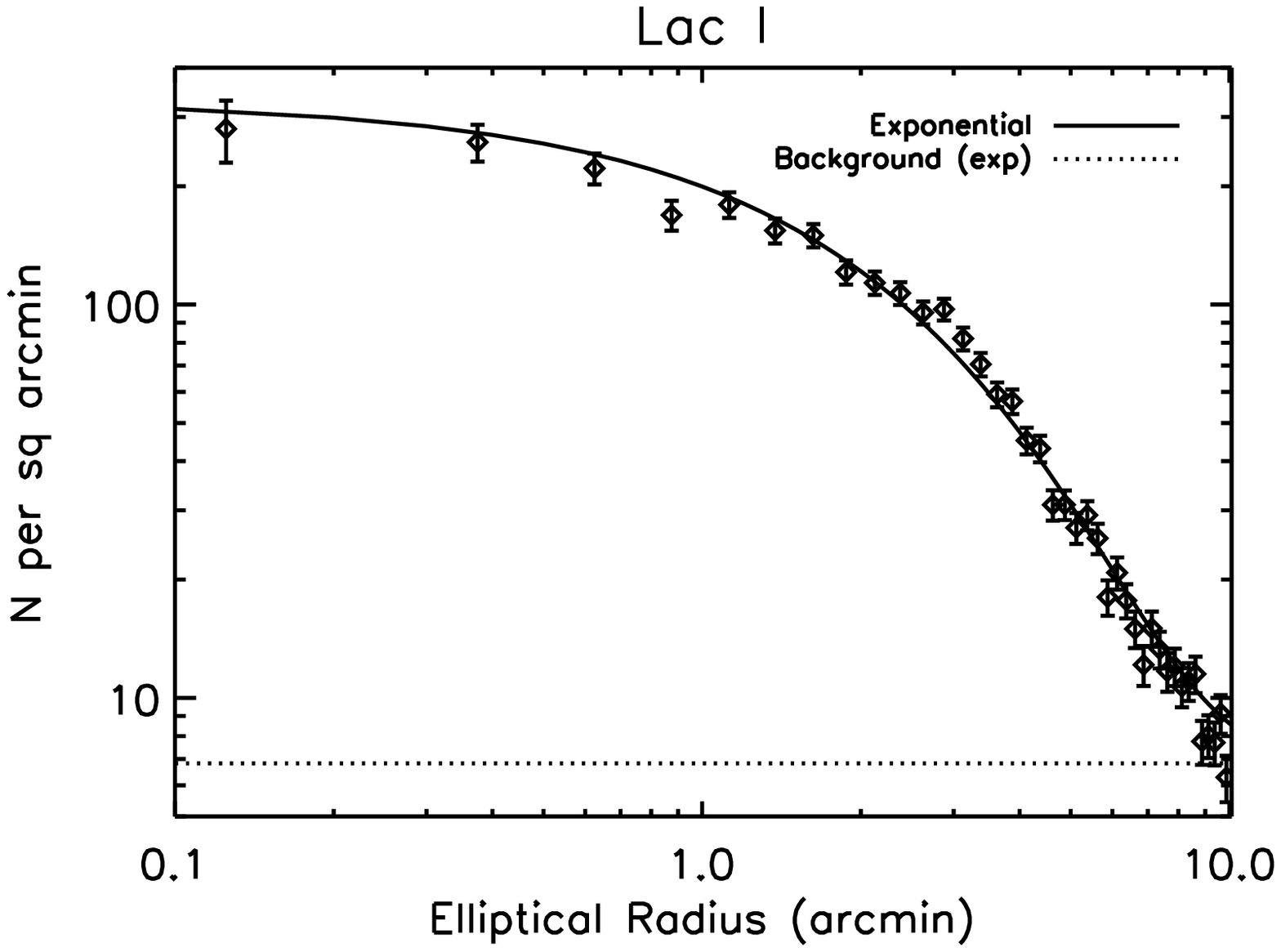}
\caption{\emph{Top left panel}. Spatial distribution of RGB stars
  across the pODI field-of-view (see selection box in the CMD
  presented in Fig.~\ref{fig:CMD}), in standard coordinates centered
  on Lac~I. The dashed ellipse is drawn at the half-light radius, the
  rectangles are the chosen field regions and the arrow indicates
  the direction towards M31. \emph{Top right panel}.  Density map of
  RGB stars in Lac~I (the grayscale key is shown on the right).
  \emph{Bottom panel}. Surface density profile of RGB stars in Lac~I
  as a function of elliptical radius. The exponential profile and
  background level, as derived from our maximum likelihood
  computation, are indicated as described in the legend. Error bars
  are Poisson errors on the number of objects in each radial bin.}
\label{fig:spat}
\end{figure*}

The first two panels of Figure~\ref{fig:CMD} show the dereddened CMD
for stars within the half-light radius of Lac~I ($r_h$ $=$
3.24$\pm$0.21~arcmin; see Section~\ref{section: structure}).  The
rightmost panel of Figure~\ref{fig:CMD} shows a rescaled CMD of the
field, for comparison. The field regions are two rectangular areas
chosen near the corners of the pODI pointing, along the dwarf galaxy's
minor axis and beyond 3$r_h$; these two regions are marked with solid
lines in the left-most panel of Figure~\ref{fig:spat}.  The total area
of the field regions is larger than the area used to produce the Lac~I
CMD, so we constructed the field CMD by randomly extracting a
proportional number of stars from the field regions so that we were
sampling equal areas of sky.  The photometric uncertainties derived
from the artificial star tests as described in the previous section
are shown in all three panels of the figure.

%\clearpage

\begin{figure}
\centering
\includegraphics[width=8.cm]{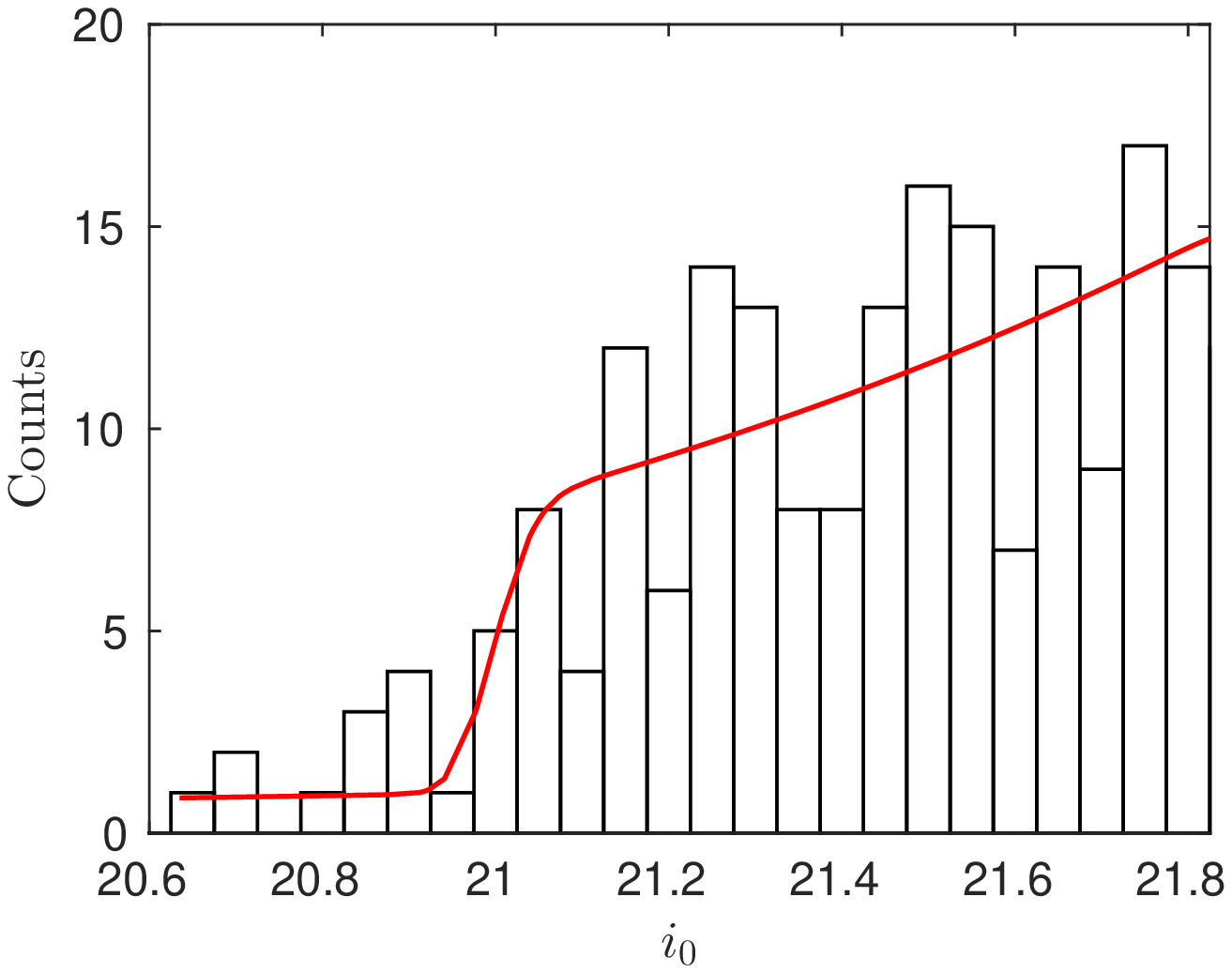}
\caption{Luminosity function of Lac~I's RGB stars within $2 r_h$ and
  with a color cut of $1.2<(g-i)_0<1.8$: the red line is the
  best-fitting model luminosity function applied in order to derive
  the TRGB luminosity.}
\label{fig:trgb}
\end{figure}

Examination of Figure~\ref{fig:CMD} shows that our follow-up
photometry from pODI reaches $\sim$3~magnitudes fainter than the
photometry presented in the Pan-STARRS1 discovery paper
\citep{martin13a}. A prominent red giant branch (RGB) is the main
feature observed in the Lac~I CMD.  The RGB coincides in location with
12~Gyr old isochrones representing a range of metallicities; see
Section~\ref{section: metallicity} for more discussion.  There is no
overdensity of bright, blue ($(g-i)_0\lesssim0.5$) sources in the
Lac~I CMD, and neither are there luminous asymptotic giant branch
(AGB) stars, which would appear just above the TRGB.  The absence of
these features suggests a predominantly old nature for Lac~I's stellar
population. Finally, the detection of a horizontal branch/red clump,
expected at $i_0\gtrsim24.5$, is hampered by the rapidly decreasing
completeness at these magnitudes (the 50\% completeness level is
marked with a dashed line in the leftmost panel of
Fig.~\ref{fig:CMD}); we are unable to draw any firm conclusions about
the presence or absence of this feature.

The top two panels of Figure~\ref{fig:spat} show the spatial
distribution of the RGB stars across the pODI field-of-view.  The
stars used to create the upper two plots in this figure are
%plotted in the left panel of
%Figure~\ref{fig:spat} 
those that appear within the RGB selection box in the leftmost panel
of Figure~\ref{fig:CMD}.  The dashed line in the left-hand panel of
Figure~\ref{fig:spat} is an ellipse centered on the galaxy that marks
the half-light radius derived in Section~\ref{section: structure}, and
the arrow
%Figure~\ref{fig:spat} 
indicates the direction toward M31. The right-hand panel of
Figure~\ref{fig:spat} shows a smoothed density map of the RGB
stars, with the surface density of stars indicated by the grayscale
value. It is clear from these two upper panels of
Figure~\ref{fig:spat} that there is an obvious overdensity of sources
that make up the main body of Lac~I, and that the galaxy has a fairly
elliptical shape. 

\begin{figure*}
\centering
\includegraphics[width=3.5in]{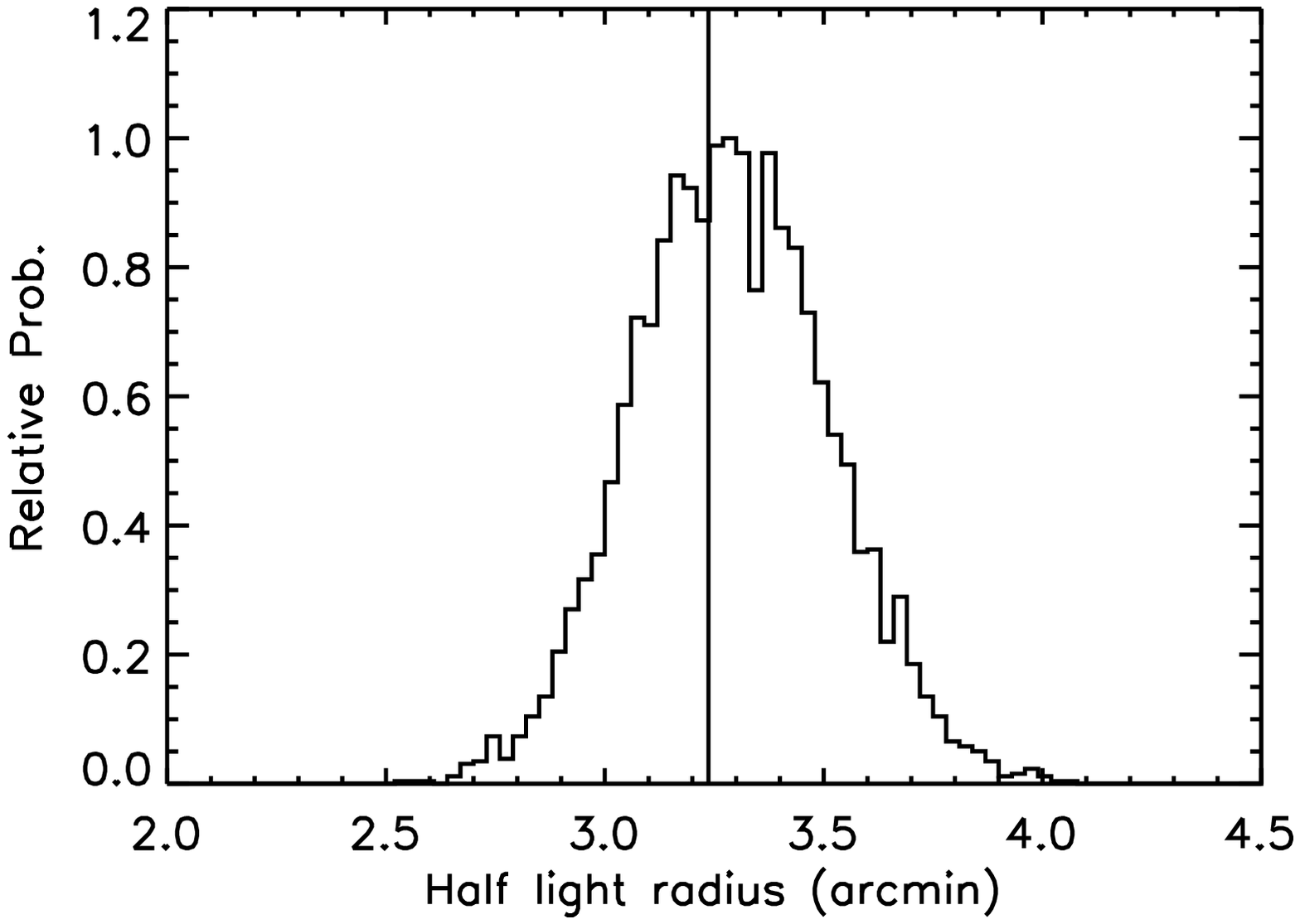}
\includegraphics[width=3.5in]{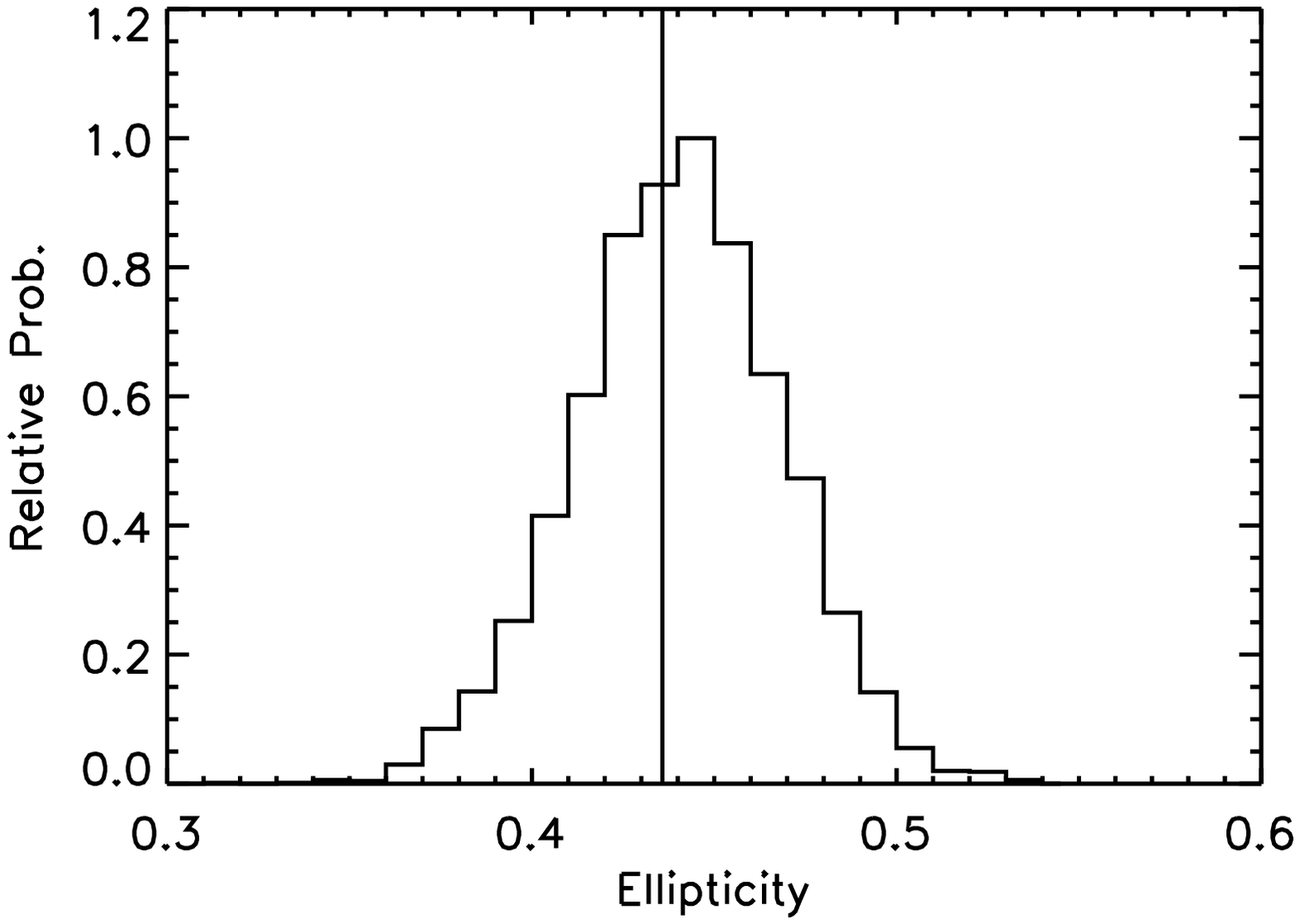}
\includegraphics[width=3.5in]{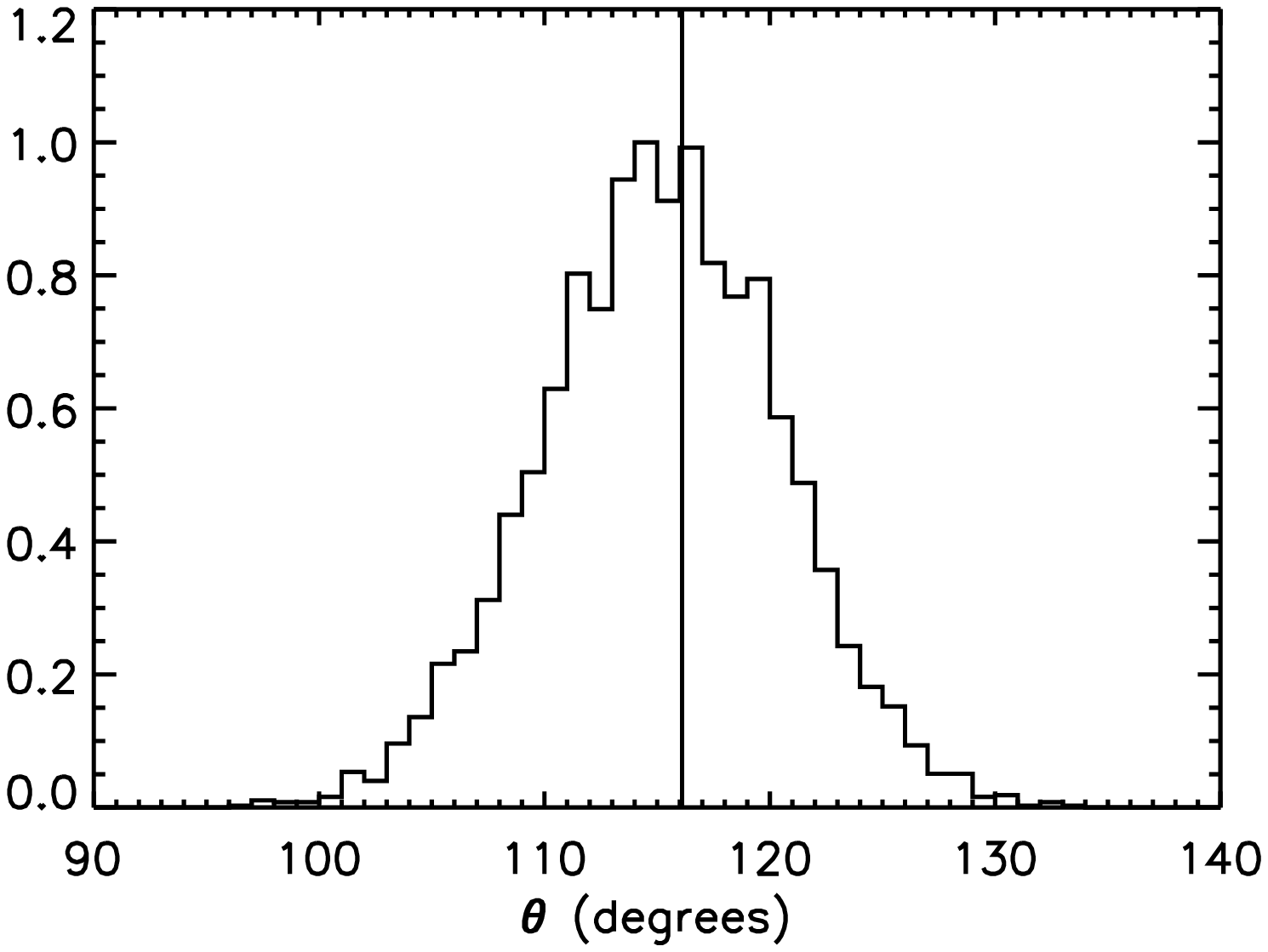}
\caption{Probability distribution functions for the structural
  parameters derived from the maximum likelihood analysis described in
  Section~\ref{section: structure}.}
\label{fig:pdf}
\end{figure*}

The distribution of contaminating objects in the CMD is deduced from
the two rectangular field regions.  As mentioned, the field regions
are located beyond $3 r_h$ (as derived in Sec.~\ref{section:
  structure}), and while we cannot rule out the presence of Lac~I
stars at these radii, their number should be negligible.
%(see right panel of Fig. \ref{fig:CMD}
%Note that the combined area of the two field regions is $\sim2.8$
%times as large as the area corresponding to the leftmost panel of
%Fig.~\ref{fig:CMD} that shows the galaxy's stellar population within
%its half-light radius.
Examination of the field CMD in the rightmost panel of
Fig. \ref{fig:CMD} illustrates that at magnitudes fainter than
$i_0\sim22.5$, unresolved background galaxies will be the main source
of contaminating objects.  At brighter magnitudes, Galactic foreground
star sequences are readily apparent.
%easily recognized. 
For all subsequent analysis in this paper, we take into account the
background/foreground contamination based on the combined CMD of the
two field regions.  The location of the RGB selection box marked in
the CMD was chosen so that it begins at the 
%repetitive, so removed this phrasing: which is marked with a solid
%red line in Fig. \ref{fig:CMD}, so that it begins at the
location of the TRGB, encompasses the full width of the RGB, and cuts
off at a faint magnitude limit of $i_0=24$ to avoid the region of the
CMD with both rapidly increasing incompleteness and photometric
uncertainties, as well as an increasing number of background galaxy
contaminants.
% chosen near the corners of the pODI pointing,
%along the dwarf galaxy's minor axis; these two regions are marked with
%solid lines in the left-hand panel of Fig. \ref{fig:spat}.  
The magnitudes and colors of the objects that appear within these two ``field''
regions of the pODI images are plotted in the rightmost panel of
Fig.~\ref{fig:CMD}. 

\subsection{TRGB Distance}
\label{section: TRGB}

The distance to Lac~I is derived by applying the TRGB method
\citep{lee93, rizzi07} to our data.  To accomplish this we adopt the
method introduced by \cite{makarov06}, modified following the approach
of \cite{wu14}.  First, a model luminosity function is convolved with
completeness, uncertainty and bias as derived from our artificial star
tests; then, we fitted this model function to Lac~I's luminosity
function (see Figure~\ref{fig:trgb}), which is derived for RGB stars
within $2 r_h$ and with a color cut of $1.2<(g-i)_0<1.8$ in order to
consider metal-poor stars and minimize contamination from foreground
Galactic stars.  This procedure determines the observed magnitude at
which the luminosity function has a sharp transition, which is
expected for old and metal-poor RGB populations at the end of this
evolutionary phase \citep[e.g.][]{salaris02}.  The absolute magnitude
for this transition is constant, and it is computed to be
$M_i=-3.44\pm0.10$ mag for the SDSS $i$-band \citep{bellazzini08}.
The uncertainty is computed with a Monte Carlo (MC) approach, by
varying the magnitude of the observed stars within the photometric
uncertainties and re-fitting the model luminosity function.  From this
analysis we derive a best-fit TRGB luminosity of $i_o=21.00\pm0.05$
mag, which translates to a distance modulus of $(m-M)_0=24.44\pm0.11$
mag, fully consistent with the $(m-M)_0=24.40\pm0.12$ mag estimate
given in the \citet{martin13a} discovery paper. The physical distance
to Lac~I is thus $773\pm40$~kpc, which is consistent with M31's
distance. Combining this with the distance to M31 and the angular
distance on the sky between the two galaxies yields a 3D distance for
Lac~I of $264\pm6$~kpc from M31.

\subsection{Structural Parameters, Luminosity, and HI Gas Content} 
\label{section: structure}

We fitted an exponential profile distribution to the surface density
profile of Lac~I, utilizing the maximum likelihood technique described
by \citet{sdssstruct}, as implemented by \citet{sand12}.  While real
satellites often have a complexity that cannot be encapsulated by
parameterized models such as the exponential profile, such profiles
are useful for quantifying the galaxies' basic structural properties
and for direct comparison with other results.  The stars selected for
the structural analysis are drawn from the RGB selection box seen in
Figure~\ref{fig:CMD}, which includes a limiting magnitude of
$i_{0}=24$ mag so as to avoid any adverse effects from crowding at
faint magnitudes.  The free parameters for our exponential profile
model are: the central position ($\alpha_0$, $\delta_0$), ellipticity
($\epsilon$, which is defined as $1-b/a$, where $b$ is the scale
length along the minor axis and $a$ is the scale length along the
major axis), 
%Note that this ellipticity definition is taken straight from Martin+08.
position angle (PA; $\theta$), half-light radius ($r_h$) and
background surface density ($\Sigma_b$).  Uncertainties on each
parameter were calculated via 1000 bootstrap resamples of the data,
from which 68\% confidence limits were derived.  The resulting
structural parameters for Lac~I are presented in
Table~\ref{tab:parms}, along with the other properties we measure for
the galaxy.  The probability distribution functions derived from
  the maximum likelihood analysis are shown in Figure~\ref{fig:pdf}.

Our structural parameters are in agreement with the discovery data of
\citet{martin13a} to within the uncertainties, although our derived
half-light radius is smaller and slightly more than 1-$\sigma$
discrepant.  The structural parameters presented here should not be
biased given the depth of the data, the stellar density contrast
compared to the background, and the field of view of the pODI data;
our data set meets all of the criteria determined by \citet{munoz12}
for deriving accurate structural parameters using our maximum
likelihood technique.

We show a one-dimensional representation of our best-fitting stellar
profile in the bottom panel of Figure~\ref{fig:spat}.  While the
binned data points are not used to fit stellar profiles -- as the
maximum likelihood technique uses the two dimensional unbinned
distribution of stars -- the agreement between the best-fitting model
stellar profile and the data points is excellent, with only minor
deviations from an exponential profile.

To measure the total magnitude of Lac I, we first sum the flux within
one half-light radius using the RGB stars within the CMD selection box
shown in Figure~\ref{fig:CMD}.  From this, we subtract off the flux
from foreground and background sources using the field regions marked
in the upper left panel of Figure~\ref{fig:spat} and rescaling to
match the area used for Lac I.  We then multiply the remaining flux by
a factor of two to account for the flux outside of Lac I's half light
radius.  We correct for stars below our detection limit by using
several metal poor ([Fe/H] = $-$1.5 to $-$2.0) and old (10-13 Gyr)
{\sc Parsec} luminosity functions (Bressan et al. 2012).  From this,
we measure $M_g$ $=$ $-$11.3$\pm$0.2 mag and $M_i$ $=$ $-$12.0$\pm$0.2
mag, and convert this to $M_V$ $=$ $-$11.4$\pm$0.3 mag using the
filter transformation of Veljanoski et al. (2013).  Our total
magnitude is consistent within the uncertainties with the value
presented in the \citet{martin13a} discovery paper ($M_V$ $=$
$-$11.7$\pm$0.7 mag), although our uncertainty is a factor of two
smaller. The central surface brightness of Lac~I is calculated using
the total magnitude and the best-fitting exponential profile described
earlier, and is $\mu_{V,0}$ $=$ 24.8$\pm$0.3 mag/arcsec$^2$. Our
measured $\mu_{V,0}$ value is a full magnitude brighter than the
corresponding value from \citet{martin13a} ($\mu_{V,0}$ $=$
25.8$\pm$0.8 mag/arcsec$^2$), although the numbers are in agreement to
within the uncertainties. As is the case with the total magnitude, the
uncertainty on our central surface brightness measurement is
substantially reduced compared to the previous measurement.  (Note
that this increase in surface brightness is not unexpected, since our
derived half-light radius is smaller.)  The derived magnitudes and
surface brightness values are included in Table~\ref{tab:parms}.

We investigated the neutral gas content of Lac~I by searching for HI
emission along the line-of-sight to the galaxy in publicly-available
Effelsberg Bonn HI Survey data (EBHIS; Winkel et al. 2016), smoothed
to a spectral resolution of 15 km~s$^{-1}$. We find no HI emission in
the smoothed spectrum at the systemic velocity for Lac~I reported by
\citet{martin14}, and compute a 5-$\sigma$, single-channel upper limit
on the HI mass within 6$'$ (1.85~$r_h$) of $M_{HI}^{lim}$ $=$ 1.9
$\times$ 10$^5 M_\sun$. This implies $M_{HI}/L_V$ $<$ 0.06
$M_\sun$/$L_\sun$ for Lac~I (see \ref{tab:parms}), and therefore that
this galaxy is gas-poor like the other dwarf spheroidal galaxies
around the Milky Way and M31 \citep{grc09,spekkens14}.

\subsection{Photometric Metallicity Distribution Functions}
\label{section: metallicity}

We derive photometric metallicities for each individual RGB star that
is located within our RGB selection box (marked in Fig. \ref{fig:CMD})
and also has $i_0 < 23$ mag.  Applying this additional faint-magnitude
cut reduces contamination significantly and keeps the photometric
uncertainties below 0.1~mag in magnitude and 0.2~mag in color, to
ensure that we are using the best-quality photometric data to yield
metallicity information. Photometric metallicities are obtained by
linearly interpolating among stellar isochrones with a fixed aged of
12~Gyr and varying metallicity ([Fe/H]$=-2.5$ to $-1.0$;
\citealt{dotter08}).  Under the assumption that the RGB width is
mainly driven by metallicity rather than age, the color of a RGB star
will correspond to an isochrone with a given metallicity (see
Fig.~\ref{fig:CMD}, and \citealt{crnojevic10} for details).  This
approximation is valid in the absence of significant intermediate-age
populations, which appears to be a safe assumption for Lac~I, based on
the appearance of the CMD.  (We do note, however, that
\citet{martin14} state that they identified ``a handful'' of carbon
stars in Lac~I based on the shape of a few of the stellar spectra they
obtained, and that the presence of these stars may suggest that Lac~I
has at least a modest intermediate-age population.)  This type of
approach provides robust results in terms of \emph{relative}
quantities such as the derivation of radial metallicity trends
\citep{vandenberg06, crnojevic14}.  We also explore the effect that a
slightly younger average age would have on the derived metallicities:
if we adopt isochrones with a fixed 8~Gyr age instead of 12~Gyr, the
metallicity values obtained for Lac~I (see below) become more
metal-rich by about 0.2~dex. (See also \citealt{crnojevic10} for an
extensive discussion on the possible effects of prolonged star
formation on the derivation of photometric metallicities.)

\begin{figure}
\centering
\includegraphics[width=3in]{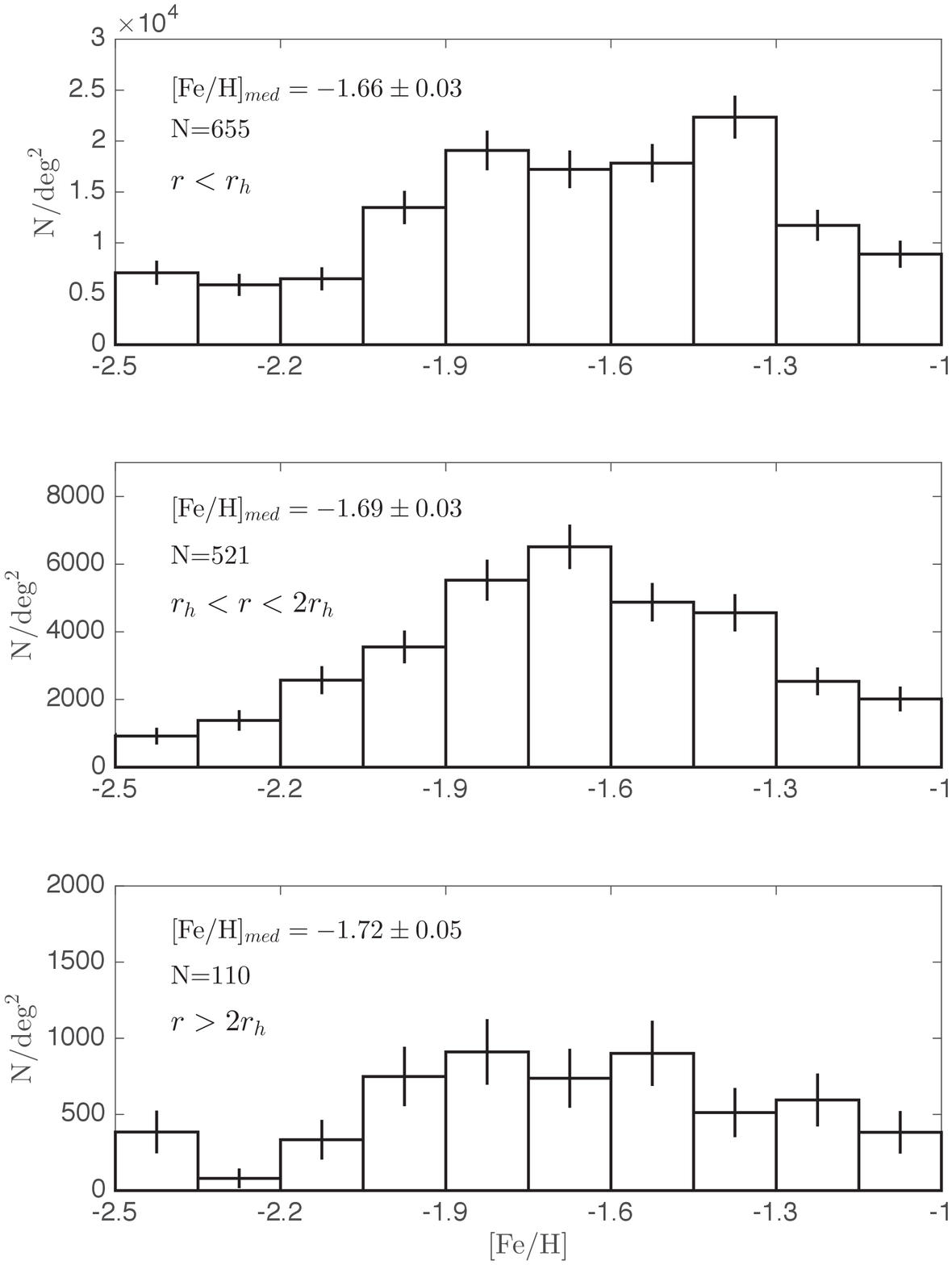}
\caption{Photometric metallicity distribution
  functions (per unit area) for stars that lie within the RGB
  selection box and have $i_0$ magnitude brighter than 23.0.  The
  three panels show MDFs for stars that meet these criteria, divided
  into three spatial regions: within one $r_h$ of the galaxy center
  (top), between one and two $r_h$ from the center, and outside of two
  $r_h$ (bottom). The number of stars per radial bin and the median
  metallicity are also reported. These MDFs have been corrected for
  contamination, as described in Section~\ref{section: metallicity}.
}
\label{fig:mdf}
\end{figure}

In Figure \ref{fig:mdf} we show the resulting metallicity distribution
functions (MDFs) for stars separated into three regions based on their
elliptical radius.  To calculate the elliptical radius ($r$) for each
star, we first transform the star's equatorial coordinates into
standard coordinates ($\xi$ and $\eta$; see Figure~\ref{fig:spat}) on
a plane tangential to the center of Lac~I.  We then take these
coordinates and use them as the $X_i$ and $Y_i$ coordinates in the
definition of elliptical radius given in \citet{sdssstruct} (their
Equation 4).  The three panels in Figure~\ref{fig:mdf} show the [Fe/H]
distribution of stars with $r < r_h$ (top panel), $r_h < r <2r_h$
(middle panel), and $r > 2r_h$ (bottom panel).  The outermost bin
includes stars out to a radius of 9~arc minutes, i.e., the largest
ellipse fully enclosed within the pODI images (which is just short of
$3r_h$).  We do not correct the derived MDFs for incompleteness, since
the RGB selection box is $\sim97\%$ complete down to $i_0<23$.  To
correct for contamination in the stellar sample, we derive an ``MDF''
for the field regions as well (strictly speaking, contaminating
sources are not RGB stars belonging to the same population, so the
derived metallicities are not meaningful), and subtract it from the
Lac~I MDFs after rescaling it to the area of each radial bin. The
percentage of contaminants in the three radial bins is $\sim2\%$,
$9\%$, and $28\%$, respectively.

We used the information about contamination gleaned from the field
``MDF'' to correct not only the Lac~I MDF data in histogram form but
also the individual photometric metallicity values of the stars in the
three radial regions shown in Figure~\ref{fig:mdf}.  We corrected the
individual photometric metallicity measurements by randomly removing
the appropriate number of stars falling within a given metallicity
range in each radial region.  In those cases when the number of
contaminants in a given metallicity and radial range was a fraction,
the value was rounded to the nearest integer, and then that number of
stars was randomly removed from the appropriate data file. 

We used the contamination-corrected MDF data to calculate the median
metallicity of the stars in the three radial ranges shown in
Figure~\ref{fig:mdf}.  Uncertainties on these median [Fe/H] values
were derived with a set of MC experiments.  First, the magnitude and
color of each star in the MDF was varied within the photometric
uncertainties for that star.  The metallicity was rederived for that
magnitude and color using the same method as for the original data
(interpolation from the isochrones).  We did this 1000 times for each
star and calculated the median [Fe/H] value for each realization, and
within each radial region.  The spread of [Fe/H] values computed from
the MC simulations yields an estimate of the uncertainty in the [Fe/H]
derived from the original data.

The median metallicities for the contamination-corrected MDFs in the
three radial regions are [Fe/H] $=$ $-1.66\pm0.03$ dex for the innermost
radial region (inside $r_h$), [Fe/H] $=$ $-1.69\pm0.03$ for stars with
$r_h < r < 2r_h$, and [Fe/H] $=$ $-1.72\pm0.05$ for stars in the outer
radial region (outside 2 $r_h$).  These values are marked in the
histogram plots in Figure~\ref{fig:mdf}.
%(An explanation of how the errors on these median values were derived
%is included later in this section.)
Combining all of the [Fe/H] values for stars within 2~$r_h$ yields a
median [Fe/H] of $-1.68$, and combining all three radial regions
yields the same median value; we have listed this value in
Table~\ref{tab:parms}. 

The median [Fe/H] values in all three of the radial regions are more
metal-rich than the [Fe/H] value of $-$2.0$\pm$0.1 derived by
\citet{martin14} from their spectroscopic observations of 126 RGB
stars in Lac~I, although in some cases the difference is not
significant when the uncertainties on these values are taken into
account.  The fixed-age approximation inherent to the photometric
metallicity derivation method can only partly explain this
discrepancy.  For the discrepancy with the \citet{martin14} data set
to be alleviated, their spectroscopic sample would need to have missed
the majority of the metal-rich subsample highlighted by our MDFs.  A
close inspection of our MDFs suggests that there could be two
  metallicity peaks, at [Fe/H]$\sim-1.8$ and [Fe/H]$\sim-1.4$
respectively, which are most pronounced in the innermost radial bin
(top panel of Fig. \ref{fig:mdf}). The distinction fades off in the
middle bin, while the last adopted bin contains few stars due to the
rapidly decreasing stellar density beyond $r=2r_h$, thus any possible
trend gets lost in the noise.  We stress that the radial MDFs have a
comparable shape when derived under the assumption of a fixed 8~Gyr
age.

We carried out the following steps to investigate whether the
  apparent multiple peaks in the MDFs might actually be statistically
  significant.  The contamination-corrected MDF data (i.e., the
  individual photometric metallicity values of the stars in the three
  radial regions shown in Figure~\ref{fig:mdf}, corrected for
  contamination) were evaluated with a series of statistical tests to
  assess the likelihood that a mixture of one, two, or three Gaussian
  distributions can produce the observed values. We used a parametric
  bootstrap Anderson-Darling (A-D) test \citep{ad52} to independently
  test each Gaussian mixture as a potential fit to the MDFs.  First, a
  set of one to three Gaussian functions was fitted to the input data,
  and best-fitting means, dispersions, and mixing proportions for each
  Gaussian were returned.  The original data were evaluated against
  the candidate mixture model using the A-D test.  Next, 1000
  synthetic data sets with the same parameters as the best-fitting
  mixture model were produced and evaluated with the A-D test. The A-D
  score derived from the observed data was then compared to the
  distribution of bootstrapped A-D scores. This testing procedure
  indicated with high confidence that the observed MDFs for all three
  radial regions are not likely to be drawn from a unimodal or bimodal
  Gaussian distribution.  The data in the outermost radial region ($r
  > 2r_h$) is also not likely to be drawn from a trimodal Gaussian
  distribution.  The tests also indicated, however, that there is a high
  probability that the MDF data for the inner two radial regions ($r <
  r_h$ and $r_h < r <2r_h$) {\it is} drawn from a mixture of three
  Gaussian distributions. The peak values (and dispersions) of the
  best-fitting Gaussian functions are [Fe/H] $=$ $-$1.3 (0.1), $-$1.8
  (0.3), and $-$2.4 (0.1) and mixing fractions of 31\%, 63\%, and 6\%,
  respectively for the stars inside the half-light radius $r_h$.  The
  corresponding values for stars with radii between $r_h$ and 2~$r_h$
  are [Fe/H] $=$ $-$1.4 (0.1), $-$1.8 (0.2), and $-$2.4 (0.1) and
  mixing fractions of 25\%, 71\%, and 5\%, respectively.

We performed additional analysis steps to determine whether the
parametric bootstrap A-D test was identifying multi-modality simply
because of the noise in the data or biases introduced during the
conversion from photometric measurements to metallicities. We
generated a synthetic MDF by populating an isochrone to produce a
12-Gyr population with [Fe/H]$\sim$$-$1.7. We convolved this catalog
with the photometric errors from our data set to obtain an
``observed'' CMD.  We derived metallicities for RGB stars within the
upper $\sim$2~magnitudes of the synthetic RGB, just as was done for
the real data. We then carried out the bootstrap testing procedure on
the synthetic [Fe/H] values and found that the best-fitting unimodal,
bimodal, and trimodal Gaussian distributions were all unlikely to
reproduce the particular distribution of the test data set at a high
confidence level (99.999\%, 98.3\%, and 99.5\% for the unimodal,
bimodal, and trimodal distributions respectively). Since the bootstrap
testing rejected not only the bimodal and trimodal mixture models but
also the single-metallicity model, we can only infer from this that
the observed trimodality that is implied by the A-D testing on the
real data may not be genuine either. Our overall conclusion based on
this analysis is that although it is possible that a dwarf galaxy with
the properties of Lac~I could indeed have multiple stellar
populations, our photometric metallicities are unable to unambiguously
demonstrate their presence or absence.

The spectroscopic metallicity value reported by \citet{martin14}
separates Lac~I from the bulk of Local Group dwarfs on a
luminosity-metallicity plot (Fig. 4 in their paper, which shows the
positions of M31 satellite dwarf galaxies on the
luminosity-metallicity (L-Z) relation derived by \citet{kirby13} from
nearby dwarfs), i.e., it is too metal-poor for its luminosity.  We can
take our updated absolute magnitude measurement, $M_V=-11.4\pm0.3$,
and our median photometric metallicity estimate of [Fe/H]
$\sim$$-$1.68 (from all of the stars in the contamination-corrected
MDFs), and compare these values to the \citet{kirby13} L-Z relation
(their Equation 3).  Assuming this relation, the predicted [Fe/H]
value for Lac~I given its luminosity is $-$1.54.  The RMS scatter of
the Kirby et al. relation is 0.16 dex, so our photometric metallicity
value is, interestingly, more in line with the expected value.

Lastly, we also looked for the presence of a metallicity gradient in
Lac~I.  We used the photometrically-derived metallicities for stars
within one and two $r_h$.  We did not correct for contamination for
this experiment 
%in this case
because the field-star contamination would be only 2\% within $r_h$
and 8\% within 2~$r_h$. The measured metallicity gradient in our data
is $d$[Fe/H]$/d(r/r_h)=0.0028\pm0.0029$ and $-0.0014\pm0.0065$ dex per
$r_h$, respectively, which is consistent with a flat metallicity
profile.  The lack of a gradient has also been observed in some dwarf
galaxies of similar luminosity beyond the Local Group
\citep[e.g.,][]{crnojevic10}, which may suggest that the galaxies are
not massive enough to favor a second episode of star formation and
thus a distinct and more metal-rich population in their central
regions.

\section{Summary and Main Conclusions}
\label{section: summary}

In this paper we presented results from deep, wide-field WIYN pODI
$g,i$ imaging 
%and photometry 
of the M31 dwarf satellite galaxy Lac~I (And~XXXI), acquired in order
to investigate the galaxy's structure, stellar populations, and
metallicity.  The CMD of this galaxy is dominated by an old ($\sim$12 Gyr)
stellar population and
%no features appear that would indicate the presence of intermediate-age stars. 
no intermediate-age stars (AGB stars) are apparent in our data. We
trace the RGB stars in the Lac~I images to a radius of $\sim$10~arcmin
($\sim$2.25~kpc) from the galaxy center and use them to derive a
distance to the galaxy as well as its structural properties. Our
measured TRGB distance of 773$\pm$40~kpc agrees with the Pan-STARRS
discovery paper distance \citep{martin13a}. Our derived 3D distance
for Lac~I to M31
%M31-centric distance
is 264$\pm$6~kpc and we 
%agrees with the Pan-STARRS discovery paper distance and confirms 
confirm the finding by \citet{martin13a} that Lac~I is in the far
western outskirts of the Andromeda galaxy halo.

Despite its relatively late discovery and distant location from its
host galaxy, Lac~I seems in other ways to be a fairly typical Local
Group dwarf spheroidal galaxy.  Our derived half-light radius $r_h$
and associated uncertainty for the galaxy, 728$\pm$47~pc, are
smaller than the Pan-STARRS1-measured value (912$^{+124}_{-93}$ pc),
although they agree within the uncertainties. \citet{brasseur11} used
the measured $V$-band absolute magnitudes and half-light radii of dSph
satellites of the Galaxy and M31
%to derive a size-luminosity relation for this class of galaxies 
and found that they follow a well-defined size-luminosity relation and
that the relations for each galaxy (the Milky Way and M31) are
statistically the same.  Given its measured $M_V$ of $-$11.4 from our
pODI data, the derived half-light radius of Lac~I is right in line
with the value predicted by the \citet{brasseur11} relation for
Andromeda dSph galaxies, 685$^{+213}_{-220}$~pc.
%Their relation for Andromeda satellite dSph galaxies is log $r_{1/2}$
%$=$ 2.35$^{+0.11}_{-0.14}$ $-$ 0.09$^{+0.02}_{-0.04}$ $x$ $(M_V + 6)$
%and yields a predicted $r_h$ value of 685 pc, and if you add the
%above errors and calculate it you get 905 pc, and if you subtract
%them you get 472 pc.
Our measured ellipticity value, $\epsilon$ $=$ 0.44$\pm$0.03 is almost
identical to the Pan-STARRS1 value, but with a formal uncertainty that
is a factor of two smaller.  This ellipticity is within the usual
range for dSph satellites of Andromeda; for the dSph galaxies in the
M31 sub-group, the range of ellipticities is 0.13$-$0.56, with a mean
$\epsilon$ of 0.31 and a dispersion of 0.12 \citep{mcconnachie12}.  We
have confirmed that this galaxy is gas-poor, with an upper limit on
the HI gas to $L_V$ ratio that is comparable to the most sensitive
limits for dSph galaxies orbiting the Milky Way \citep{spekkens14}.

Our investigation of the MDF of Lac~I -- which we derived from
photometry of the RGB stars with $i_0$ $<$ 23 mag --- indicates that
this galaxy is metal-poor, with a median [Fe/H] of $-$1.68$\pm$0.03
for stars within 2~$r_h$, but a broad range of stellar metallicities,
from [Fe/H]$\sim$ $-1$ to $-2.5$.  Although the appearance of the MDFs
in certain radial regions seems to suggest the presence of multiple
metallicity peaks, statistical testing performed on the distributions
indicates that the multi-modality may not be real.  We find no
evidence for a radial gradient in our photometric metallicities.
Lastly, we note that the median metallicity derived from our RGB star
photometry is higher than the median metallicity of [Fe/H] $=$
$-2.0$$\pm$0.1 measured by \citet{martin14} from their spectroscopic
observations of 126 bright ($i_{P1,0}$ \lapp 20.5) RGB stars in Lac~I,
and the somewhat higher metallicity we measure moves Lac~I closer to
the expected $L-Z$ relation for dwarf galaxies \citep{kirby13}. The
causes of the differences in the photometric versus spectroscopic
metallicities are not obvious and seem to warrant further
investigation.

\acknowledgments 

We are grateful to the staff of the WIYN Observatory and Kitt Peak
National Observatory for their assistance with using WIYN and the pODI
camera to obtain the data used for this study.  We thank the staff
members at WIYN, NOAO, and Indiana University Pervasive Technology
Institute who designed, implemented, and helped us work with the
ODI-PPA and produce high-quality science results with our pODI
images. We also thank the anonymous referee for helpful comments that
improved the paper.  K.L.R. was supported by NSF Faculty Early Career
Development (CAREER) award AST-0847109 or by NSF Astronomy \&
Astrophysics Research Grant number AST-1615483 during the time when
this research was carried out.  D.J.S. acknowledges support from NSF
Astronomy \& Astrophysics Research Grant number AST-1412504.  S.J.
acknowledges support from the Australian Research Council's Discovery
Project funding scheme (DP150101734).  K.S. acknowledges support from
the Natural Sciences and Engineering Research Council of Canada
(NSERC).  This research has made use of the NASA/IPAC Extragalactic
Database (NED) which is operated by the Jet Propulsion Laboratory,
California Institute of Technology, under contract with the National
Aeronautics and Space Administration.

\software{ODI-PPA (Gopu et al. 2014), QuickReduce pipeline (Kotulla
  2014), DAOPHOT (Stetson 1987), ALLFRAME (Stetson 1994)}

{}

%\clearpage

\begin{deluxetable}{lccccrccc}
%\tablewidth{200pt}
%\tabletypesize{\scriptsize}
%\tabletypesize{\small}
\tablecaption{Photometry of Resolved Stars in the WIYN pODI Images of Lac~I \label{tab:photometry}}
\tablehead{
\colhead{\#}& \colhead{$\alpha$ (2000)}& \colhead{$\delta$ (2000)}&
\colhead{$g_o$}& \colhead{$\sigma_{g}$}& \colhead{$A_g$}
& \colhead{$i_o$}&\colhead{$\sigma_{i}$} & \colhead{$A_i$}\\
\colhead{}& \colhead{(deg)}& \colhead{(deg)}&
\colhead{(mag)}& \colhead{(mag)}& \colhead{(mag)}
&\colhead{(mag)}&\colhead{(mag)}&\colhead{(mag)} 
}
\startdata
1 & 344.37250 & +41.458038 & 16.73 & 0.01 & 0.52 & 15.64 & 0.01 & 0.27\\
2 & 344.37418 & +41.425993 & 25.58 & 0.29 & 0.53 & 24.53 & 0.16 & 0.27\\
3 & 344.37427 & +41.425386 & 24.87 & 0.12 & 0.53 & 25.14 & 0.20 & 0.27\\
4 & 344.37434 & +41.436219 & 24.95 & 0.11 & 0.53 & 24.22 & 0.08 & 0.27\\
5 & 344.37458 & +41.422852 & 23.43 & 0.04 & 0.53 & 23.08 & 0.04 & 0.27\\
6 & 344.37478 & +41.428685 & 26.29 & 0.39 & 0.53 & 25.23 & 0.25 & 0.27\\
7 & 344.37479 & +41.410686 & 26.32 & 0.40 & 0.53 & 25.55 & 0.27 & 0.27\\
8 & 344.37481 & +41.397515 & 22.83 & 0.02 & 0.54 & 20.52 & 0.01 & 0.28\\
9 & 344.37486 & +41.430424 & 24.32 & 0.07 & 0.53 & 23.45 & 0.04 & 0.27\\
10 & 344.37489 & +41.383481 & 25.37 & 0.26 & 0.54 & 25.27 & 0.31 & 0.28\\

\enddata
\tablecomments{This table is available in its entirety in a machine-readable form in
the online journal. A small portion of the data is shown as an example
of the form and content of the table.}
\end{deluxetable}

\begin{deluxetable}{lr} 
%\tablewidth{0pt}
%\tablewidth{420pt}
%\tabletypesize{\scriptsize}
\tablecaption{Properties of Lac~I \label{tab:parms}}
\tablehead{
\colhead{Property}& \colhead{Value}  
}
\startdata
\hline\\
%Right Ascension ($\alpha$, $^{\circ}$)	 & 344.55527 $\pm$ 0.00078\\ 	
%Declination ($\delta$, $^{\circ}$)             & 41.298206 $\pm$ 0.000583  \\ 	
Right Ascension ($\alpha$)	 & 22h 58m 13.26s $\pm$ 0.19s\\
Declination ($\delta$)             & 41$^{\circ}$17$'$53.5$''$ $\pm$ 2.1$''$\\ 	
$M_V$ (mag)	 	& $-$11.4 $\pm$ 0.3 \\
$M_g$ (mag) 	 	& $-$11.3 $\pm$ 0.2 \\
%$M_g$=$-$11.3$\pm$0.2
%and $M_i$=$-$12.0$\pm$0.2 mag
%so I derived g-i = 0.7+/-0.3
($g$$-$$i$)$_o$ (mag) & 0.7 $\pm$ 0.3 \\
$(m-M)_0$ (mag)  & 24.44 $\pm$ 0.11 \\
$D$ (kpc) & 773 $\pm$ 40 \\
$D_{\rm M31}$ (kpc) & 264 $\pm$  6 \\
$\mu_{V,0}$ (mag/arcsec$^2$) & 24.8 $\pm$ 0.3\\ 
$r_h$ (arcmin) &  3.24 $\pm$ 0.21 \\
$r_h$ (pc) &  728 $\pm$ 47 \\
Ellipticity ($\epsilon$)& 0.44 $\pm$ 0.03\\
Position angle ($\theta$, $^{\circ}$ East of North) & 116.1 $\pm$ 5.4\\
$M_{HI}/L_V$ ($M_\sun$/$L_\sun$) & $<$0.06\\
%${\rm [Fe/H]}_{\rm med}$ inside $2~r_h$ (dex) & $-1.68\pm0.03$ \\
${\rm [Fe/H]}_{\rm med}$ (dex) & $-1.68\pm0.03$ \\
%\hline
\enddata
\end{deluxetable}


\begin{thebibliography}{}
% example bibitem

\bibitem[Adams et al. (2013)]{adams13} Adams, E.A.K., Giovanelli, R.,
  \& Haynes, M.P. 2013, \apj, 768, 77

\bibitem[Ahn et al. (2012)]{ahn12} Ahn, C., Alexandroff, R., Allende
  Preito, C., et al. 2012, ApJS, 203, 21

\bibitem[Anderson \& Darling (1952)]{ad52} Anderson, T. \& Darling,
  D. 1952, Ann. Math. Statist. 23, No. 2, pp. 193$-$212

\bibitem[Bechtol et al. (2015)]{bechtol15} Bechtol, K., et al. 2015,
  \apj, 807, 50

\bibitem[Bell et al. (2011)]{bell11} Bell, E.F., Slater, C.T., \&
  Martin, N.F. 2011, \apj, 742, L15

\bibitem[Bellazzini (2008)]{bellazzini08} Bellazzini, M. 2008, MmSAI,
  79, 440

\bibitem[Belokurov et al. (2007)]{belokurov07} Belokurov, V. et
  al. 2007, \apj, 654, 897

\bibitem[Brasseur et al. (2011)]{brasseur11} Brasseur, C.M., Martin,
  N.F., Maccio, A.V., Rix, H.-W., \& Kang, X. 2011, \apj, 743, 179 

\bibitem[Bullock \& Johnston (2005)]{bullock05} Bullock, J.S. \&
  Johnston, K.V. 2005, \apj, 635, 931

\bibitem[Bressan et al. (2012)]{Bressan12} Bressan, A., Marigo, P.,
  Girardi, L. {et~al.} 2012, MNRAS, 427, 127 

\bibitem[Crnojevi\'c et al. (2010)]{crnojevic10} Crnojevi\'c, D.,
  Grebel, E.K., \& Koch, A. 2010, A\&A, 516, 85

\bibitem[Crnojevi\'c et al. (2014)]{crnojevic14} Crnojevi\'c, D.,
  Ferguson, A.M.N., Irwin, M.J., McConnachie, A.W., Bernard, E.J.,
  Fardal, M.A., Ibata, R.A., Lewis, G.F., Martin, N.F., Navarro, J.F.,
  Noel, N.E.D., \& Pasetto, S. 2014, \mnras, 445, 3862

\bibitem[Dotter et al. (2008)]{dotter08} Dotter, A., Chaboyer, B.,
  Jevremovi\'c, D., Kostov, V., Baron, E., \& Ferguson, J.W. 2008, ApJS,
  178, 89

\bibitem[Grcevich \& Putman (2009)]{grc09} Grcevick, J. \& Putman,
  M.E. 2009, ApJ, 696, 385

\bibitem[Gopu et al. (2014)]{gopu14} Gopu, A., Hayashi, S., Young,
  M.D., Harbeck, D.R., Boroson, T., Liu, W., Kotulla, R., Shaw, R.,
  Henschel, R., Rajagopal, J., Stobie, E., Knezek, P., Martin, R.P.,
  Archbold, K. 2014, Proceedings of the SPIE, Vol. 9152, p. 91520E

\bibitem[Harbeck et al. (2014)]{harbeck14} Harbeck, D.R., Boroson, T.,
  Lesser, M., Rajagopal, J., Yeatts, A., Corson, C., Liu, W.,
  Dell'Antonio, I., Kotulla, R., Ouellette, D., Hooper, E., Smith, M.,
  Bredthauer, R., Martin, P., Muller, G., Knezek, P., \& Hunten,
  M. 2014, Proceedings of the SPIE, Vol. 9147, p. 9147E

\bibitem[Ibata et al. (2013)]{ibata13} Ibata, R.A., Lewis, G.F., Conn,
  A.R., Irwin, M.J., McConnachie, A.W., Chapman, S.C., Collins, M.L.,
  Fardal, M., Ferguson, A.M.N., Ibata, N.G., Mackey, A.D., Martin,
  N.F., Navarro, J., Rich, R.M., Valls-Gabaud, D., Widrow, L.M. 2013,
  Nature, 493, 62

\bibitem[Janesh et al. (2015)]{janesh15} Janesh, W.F., Rhode, K.L.,
  Salzer, J.J., Janowiecki, S., Adams, E.A.K., Haynes, M.P.,
  Giovanelli, R., Cannon, J.M., \& Munoz, R.R. 2015, \apj, 811, 35

\bibitem[Jenkins et al. (2001)]{jenkins01} Jenkins, A., Frenk, C.S.,
  White, S.D.M., Colberg, J.M., Cole, S., Evrard, A. E., Couchman,
  H.M.P., Yoshida, N. 2001, \mnras, 321, 372

\bibitem[Kim et al. (2015)]{kim15} Kim, D., Jerjen, H., Mackey, D.,
  Da~Costa, G.S., Milone, A.P. 2015, \apj, 804, L44

\bibitem[Kirby et al. (2013)]{kirby13} Kirby, E., Cohen, J.,
  Guhathakurta, P., Cheng, L., Bullock, J.S., \& Gallazzi, A. 2013,
  \apj, 779, 102

\bibitem[Klypin et al. (1999)]{klypin99} Klypin, A., Kravtsov, A. V.,
  Valenzuela, O., \& Prada, F. 1999, \apj, 522, 82 

\bibitem[Koposov et al. (2015)]{koposov15} Koposov, S.E., Belokurov,
  V., Torrealba, G., \& Evans, N.W. 2015, \apj, 805, 130

\bibitem[Kotulla (2014)]{kotulla14} Kotulla, R. 2014, in Astronomical
  Data Analysis Software and Systems XXIII, Astronomical Society of
  the Pacific Conference Series, Vol. 485, eds. N. Manset \&
  P. Forshay, 375

\bibitem[Lee et al. (1993)]{lee93} Lee, M.G., Freedman, W.L., \&
  Madore, B.F. 1993, \apj, 417, L553

\bibitem[Makarov et al. (2006)]{makarov06} Makarov, D., Makarova, L., 
  Rizzi, L., Tully, R.B., Dolphin, A.E., Sakai, S., \& Shaya, E.J. 
  2006, AJ, 132, 2729

\bibitem[Martin et al. (2006)]{martin06} Martin, N., et al. 2006,
  \mnras, 371, 1983

\bibitem[{{Martin} {et~al.}(2008){Martin}, {de Jong}, \& {Rix}}]{sdssstruct}
{Martin}, N.~F., {de Jong}, J.~T.~A., \& {Rix}, H.-W. 2008, \apj, 684, 1075

\bibitem[Martin et al. (2013a)]{martin13a} Martin, N.F., Slater, C.T.,
  Schlafly, E.F., Morganson, E., Rix, H.-W., Bell, E.F., Laevens,
  B.P.M., Bernard, E.J., Ferguson, A.M.N., Finkbeiner, D.P., Burgett,
  W.S., Chambers, K.C., Hodapp, K.W., Kaiser, N., Kudritzki, R.-P.,
  Magnier, E.A., Morgan, J.S., Price, P.A., Tonry, J.L., \& Wainscoat,
  R.J. 2013, \apj, 772, 15

\bibitem[Martin et al. (2013b)]{martin13b} Martin, N.F., Schlafly,
E.F., Slater, C.T., Bernard, E.J., Rix, H.-W., Bell, E.F., Ferguson,
A.M.N., Finkbeiner, D.P., Laevens, B.P.M., Burgett, W.S., Chambers,
K.C., Draper, P.W., Hodapp, K.W., Kaiser, N., Kudritzki, R.-P.,
Magnier, E.A., Metcalfe, N., Morgan, J.S., Price, P.A., Tonry, J.L.,
Wainscoat, R.J., \& Waters, C. 2013, \apj, 779, L10

\bibitem[Martin et al. (2014)]{martin14} Martin, N.F., Chambers, K.C.,
  Collins, M.L.M., Ibata, R.A., Rich, R.M., Bell, E.F., Bernard, E.J.,
  Ferguson, A.M.N., Flewelling, H., Kaiser, N., Magnier, E., Tonry,
  J.L., \& Wainscoat, R.J. 2014, \apj, 793, L14

\bibitem[McConnachie et al. (2009)]{mcconnachie09} McConnachie, A.W.,
  et al. 2009, Nature, 461, 66

\bibitem[McConnachie (2012)]{mcconnachie12} McConnachie, A.W.,
  2012, AJ, 144, 4

\bibitem[Moore et al. (1999)]{moore99} Moore, B., Ghigna, S.,
  Governato, F., Lake, G., Quinn, T., Stadel, J., \& Tozzi, P. 1999,
  \apj, 524, L19

\bibitem[Mu{\~n}oz et al. (2012)]{munoz12} Mu{\~n}oz, R.R.,
  Padmanabhan, N., \& Geha, M. 2012, \apj, 745, 127

\bibitem[Rizzi et al. (2007)]{rizzi07} Rizzi, L., Tully, R.B.,
  Makarov, D., Makarova, L., Dolphin, A.E., Sakai, S., \& Shaya,
  E.J. 2007, \apj, 661, 815

\bibitem[Salaris et al. (2002)]{salaris02} Salaris, M., Cassisi, S.,
  \& Weiss, A. 2002, PASP, 114, 375

\bibitem[Sand et al. (2012)]{sand12} Sand, D.J., Strader, J., Willman,
  B., Zaritsky, D., McLeod, B., Caldwell, N., Seth, A., \& Olszewski,
  E. 2012, \apj, 756, 79

\bibitem[Schlafly \& Finkbeiner (2011)]{schlafly11} Schlafly,
  E.F. \& Finkbeiner, D.P. 2011, \apj, 737, 103

\bibitem[Schlegel et al. (1998)] {schlegel98} Schlegel, D.J.,
  Finkbeiner, D.P., \& Davis, M. 1998, \apj, 500, 525

\bibitem[Simon \& Geha (2007)]{simon07} Simon, J.D. \& Geha, M. 2007,
  \apj, 670, 313

\bibitem[Spekkens et al. (2014)]{spekkens14} Spekkens, K., Urbancic,
  N., Mason, B.S., Willman, B., \& Aguirre, J.E. 2014, ApJ, 795, L5

\bibitem[Stetson (1987)]{stetson87} Stetson, P.B. 1987, PASP, 99, 191

\bibitem[Stetson (1994)]{stetson94} Stetson, P.B. 1994, PASP, 106, 250

\bibitem[VandenBerg et al. (2006)]{vandenberg06} VandenBerg, D.A.,
  Bergbusch, P.A., \& Dowler, P.D. 2006, ApJS, 162, 375

\bibitem[Veljanoski et al. (2013)]{Veljanoski13} Veljanoski, J.,
  Ferguson, A. M., Huxor, A.P. {et~al.} 2013, MNRAS, 435, 3654 

\bibitem[Willman et al. (2005)]{willman05} Willman, B. et al. 2005,
  \apj, 626, L85

\bibitem[Winkel et al. (2016)]{winkel16} Winkel, B., Kerp, J., Floer,
  L., Kalberla, P.M.W., Ben Bekhti, N., Keller, R., \& Lenz, D. 2016,
  A\&A, 585, 41

\bibitem[Wu et al. (2014)]{wu14} 	
	Wu, P., Tully, R.B., Rizzi, L., Dolphin, A.E., Jacobs, B.A.,
  \& Karachentsev, I.D. 2014, AJ, 148, 7

\end{thebibliography}
\end{document}